\documentclass[iop]{emulateapj}

\usepackage{natbib}
\usepackage{epsfig}

\usepackage{apjfonts}

\newcommand{\figref}[1]{Figure \ref{#1}}
\newcommand{\eqref}[1]{equation (\ref{#1})}

\newcommand{\sub}[1]{\ensuremath{_{\mbox{\scriptsize#1}}}}

\begin{document}
\slugcomment{To Appear in the Astrophysical Journal}
\title{Gaps in Protoplanetary Disks as Signatures of Planets:
I.~Methodology and Validation}
\author{Hannah Jang-Condell}
\affil{Department of Physics \& Astronomy, University of Wyoming, 
Laramie, WY 82071, U.S.A.}
\author{Neal J.~Turner}
\affil{Jet Propulsion Laboratory, California Institute of Technology, 
Pasadena, CA 91109, U.S.A.}
\shorttitle{\sc Gaps in Protoplanetary Disks}
\shortauthors{\sc Jang-Condell \& Turner}

\begin{abstract}
We examine the observational consequences of partial gaps being 
opened by planets in protoplanetary disks.  We model the disk 
using a static $\alpha$-disk model with detailed radiative 
transfer, parametrizing the shape and size of the partially cleared gaps 
based on the results of hydrodynamic simulations.  
Shadowing and illumination by stellar irradiation 
at the surface of the gap leads to increased contrast as the 
gap trough is deepened by shadowing and cooling and the far 
gap wall is puffed up by illumination and heating.  
In calculating observables, we find that multiple scattering is important 
and derive an approximation to include these effects.  
A gap produced by a 200 $M_{\earth}$ (70 $M_{\earth}$)
planet at 10 AU can lower/raise the 
midplane temperature of the disk by up to 
$\sim-25/+29\%$ ($\sim-11/+19$) by shadowing 
in the gap trough and illumination on the far shoulder of the gap.  
At the distance of Taurus, this gap would be resolvable with 
$\sim0.01\arcsec$ angular resolution.  
The gap contrast is most significant in scattered light and at 
thermal continuum wavelengths characteristic of the surface temperature, 
reducing or raising the surface brightness by up to order of magnitude. 
Since gaps sizes are correlated to planet mass, this is a promising 
way of finding and determining the masses of planets embedded 
in protoplanetary disks.  
\end{abstract}


\section{Introduction}

With the advent of adaptive optics systems on large ground-based 
optical and near-IR telescopes, 
we are beginning to be able to image protoplanetary 
disks around young stars \citep[e.g.~][LkCa 15]{2010Thalmann_etal}.    
These gas-rich disks are where we expect giant planets to form, 
since gas giants need that large reservoir of gas from which 
to accrete their massive envelopes.  However, the 
interpretation of structure in these imaged disks can be problematic,
because scattered light traces only the optically thin 
and diffuse surface layers of the disks rather than 
the overall structure of the disk \citep{2007HJCBoss}.  
On the other hand, if growing planets do indeed significantly 
perturb the disks in which they are born, observation of these 
signatures can mean the detection of nascent planets 
in protoplanetary disks.  

\citet[][henceforth JC08]{HJC_model} calculated the shape of a 
dimple created by 
the gravitational potential of a planet embedded in a disk and 
the accompanying thermal perturbations in shadowed and illuminated 
regions of the dimple.  
\citet[][henceforth JC09]{2009HJC} predicted the 
observability of these dimples.  Neither of the papers considered 
the clearing of an annular gap in the disk by tidal forces.  
Planets above about 30 M$_{\oplus}$ are able to open 
partial gaps in disks, as demonstrated in numerical 
hydrodynamic simulations \citep[e.g.][]{bate}.  
These gaps are much larger in scale than the local 
dimples modeled in JC08 and JC09 and are therefore a 
promising way of detecting and characterizing 
planets in protoplanetary disks.  
If the gap is well resolved enough to determine its 
depth and width, then we can determine the mass of the planet 
to within a factors of a few.  

JC09 considered only single scattering and direct thermal emission 
to calculate observable signatures of disk perturbations.  
However, multiple scattering, particularly at high albedos, 
can significantly increase disk brightnesses.  
While Monte Carlo methods directly calculate multiple scattering, 
they are computationally intensive so that iteratively calculating 
the disk structure is prohibitively time consuming
\citep[e.g.~][]{2004MNRAS.351..607W,2004DullemondDominik,2008ApJ...689..513T,2008Pinte_etal,2010MuldersDominikMin}.
The approach here is to exploit a few analytic approximations to 
the solutions of the radiative transfer equations to efficiently 
calculate the disk structure.  

Numerous people have carried out hydrodynamic simulations of 
gap opening in disks by planets 
\citep[e.g.][]{2008PaardekooperPapaloizou,2009AyliffeBate,2006deValBorro_etal,bate,2008MNRAS.387..387E}
However, the effects of illumination of these gaps 
by the central star has not been well-studied.  
\citet{2002Wolf_etal} examine this problem in a flat 
disk with constant $H/r=0.05$ and find that the far edge 
of the gap is not illuminated.  On the other hand, 
\citet{2006Varniere_etal} find that the far edge in a flared 
disk, which is more typical of T Tauri disks, 
is heated and puffed up, creating a positive feedback loop 
that enhances the appearance of the gap.  
In both these cases, the gap was created by a Jupiter-mass planet, 
which clears nearly all the material from its orbital, creating 
a deep and wide gap.  In this work, we focus on partially cleared 
gaps created by sub-Jupiter mass planets.  

A full three-dimensional hydrodynamic simulation including 
radiative transfer is very computationally 
intensive, so we simplify the problem by assuming a fixed 
disk structure in terms of its radial surface density profile, 
$\Sigma(r)$.  
To further simplify the computation, we assume an axisymmetric disk 
structure.  This means that we do not include 
the behavior of spiral density waves, 
but only the large scale gap opened by the planet.  

\section{The JC Model}

The methods used for calculating radiative transfer in protoplanetary 
disks for both radiative heating and observable quantities 
are those used in JC08 and JC09.  In this work, we now apply these 
methods to an axisymmetric disk structure rather than an 
azimuthally limited local perturbation, and we 
account for multiple scattering in the observables.  
We will refer to these methods as the JC model, 
to constrast to Monte Carlo methods.  

The disk model incorporates radiative transfer based on the methods of 
\citet{paper1} and \citet{paper2} and adopted to self-consistently 
determine the disks thermal and pressure structure as described in 
JC08\@.  Stellar irradiation is an important heating source 
in protoplanetary disks, and the amount of heating is sensitive to the 
angle of incidence at the disk surface.  If the shape of the surface 
of the disk changes on scales smaller than the disk thickness, 
the plane-parallel approximation fails.  To account for this kind 
of disk structure, radiative heating is integrated piecewise 
over the surface of the disk rather than assuming a thin 
plane-parallel disk model.  This method of calculating radiative transfer 
is fully-three dimensional and not directionally dependent on a 
fixed coordinate system.  More detailed explanation of the method
can be found in JC08 and references therein.  
Calculation of the observables is done as detailed in JC09,
with modifications to approximate the effects of multiple scattering,
as discussed below.  
In this section, we summarize the essential features of the model.  

\subsection{Initial Conditions}\label{initcond}

We follow the same method as used in JC08 and JC09
to calculate the initial disk structure.  
The stellar parameters are 
mass $M_*=1\,M_{\odot}$, 
radius $R_*=2.6\,R_{\odot}$, 
and effective temperature $T\sub{eff}=4280$ K, 
consistent with a protostar with an age of 1 Myr \citep{siess_etal}.  
We assume a constant-$\alpha$ disk model where the 
viscosity is parameterized as $\nu=\alpha\sub{ss} c_s H$
\citep{shaksun}.  
The disk parameters are 
accretion rate $\dot{M}=10^{-8}\,M_{\sun}\,\mbox{yr}^{-1}$ and 
viscosity parameter $\alpha\sub{ss}=0.01$, 
parameters typical for T Tauri stars.  

The initial conditions are calculated in a two-step process.
In both steps, we iteratively calculate the density 
and temperature structure of the disk, including 
radiative transfer of the stellar irradiation at the disk 
surface.  Viscous heating is included, but is only important 
inwards of a few AU\@.  Beyond this distance, 
the primary heating source is stellar irradiation, which depends sensitively
on the angle of incidence at the surface of the disk. 
The surface is defined to be where the Planck-averaged 
optical depth to stellar irradiation is 2/3. 

In step 1, 
we generate a locally plane-parallel model for the entire disk.  
For this initial locally plane-parallel model, we 
calculate the disk at logarithmically spaced 
intervals of $\sqrt{2}$ from 0.25 AU to 256 AU\@.  

In step 2,
we select a radially and azimuthally limited wedge of this 
disk to calculate in more detail, including the three dimensional 
curvature of the wedge, but assuming axisymmetry.  
The model presented here is 
a slice with radial range ($r$) from 3 to 20.8 AU and 
vertical distance from the midplane ($z$) from 0 to 5.5 AU 
in order to model the gap created by at planet at 10 AU\@. 
We sample the initial density and temperature 
at 100 grid points in both $r$ and $z$ directions, 
using cylindrical coordinates.  
In order to ensure that that this initial disk is in
thermal and hydrostatic equilibrium, we recalculate the 
heating from stellar irradiation, now removing the 
assumption that the disk is locally plane parallel 
and including the full three-dimensional curvature of the disk.  
The resulting density profile is $\Sigma_0(r)$.  
The total disk mass within the simulation boundaries is 0.0025 $M_{\sun}$.  

\subsection{Gap Modeling}

The depth of a gap that can be opened by a planet 
depends not only on the mass ratio between planet and star, 
but also on the disk properties.  
In a low viscosity disk, 
the critical threshold for gap-opening is generally considered 
to be the mass at which the planet's 
Hill radius equals the thermal scale height of the disk.  
The Hill radius is 
$r\sub{Hill} = (q/3)^{1/3} a$
where $q\equiv M_p/M_*$ is the mass ratio of the planet to the star. 
The thermal scale height is 
$H = c_s / \Omega_K$ where 
$c_s\equiv\sqrt{kT/\mu}$ is the thermal sound speed 
and $\Omega_K\equiv\sqrt{GM_*/a^3}$  is the Keplerian orbital angular speed.  
Here, $k$ is Boltzmann's constant; 
$T$ is the disk's midplane temperature;
$\mu$ is the mean molecular weight of 
of the gas, assumed to be primarily molecular hydrogen; 
and $G$ is Newton's gravitational constant.  

The gap opening criterion for a viscous disk has been 
found empirically to be 
\begin{equation}\label{eq:viscgapcrit}
\frac{3}{4}\frac{H}{r\sub{Hill}} + \frac{50}{q {\cal R}}
\lesssim 1
\end{equation}
where Reynolds number is defined as ${\cal R}\equiv r^2\Omega_P/\nu$
\citep{2006CridaMorbidelliMasset}. 
Their results are also consistent with those of 
\citet{2004ApJ...612.1152V}.  

For the purposes of this paper, we adopt gap profiles 
similar to those calculated by \citet{bate}, who model 
gap opening by planets of varying masses in a 
disk using three-dimensional hydrodynamic simulations, 
focusing specifically on planets that only partially 
open gaps in disks.  
A gap is modeled as an ad hoc perturbation imposed on the initial 
conditions.  The surface density of a disk modified by a gap 
of width $w$, depth $d$, and position $a$ is
\begin{equation}\label{eq:gap}
\label{gapdenprof}
\Sigma(r) = \Sigma_0(r) \left\{1-d\exp[-(r-a)^2/(2w^2)]\right\}.
\end{equation}
This is consistent with the results of 
\citep{2007MNRAS.377.1324C}, who find that planets 
whose masses are less than that of the disk open local 
gaps in disks rather than clearing inner cavities.  
Since the study presented here addresses planets under 
a Jupiter mass ($M_J$), they are well under the disk mass. 

\begin{figure}[htbp]
\plotone{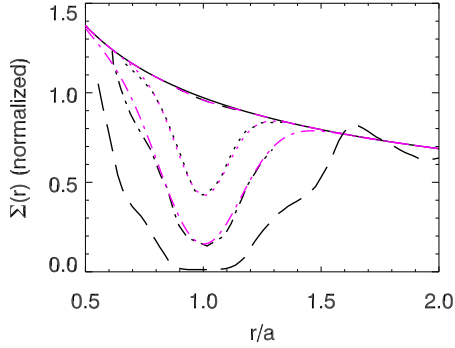}
\caption{\label{bategapfit}
Profiles of gaps carved by planets in disks and the Gaussian fits to them.  
The unperturbed disk profile is plotted as a solid line, 
while the dashed, dotted, dot-dashed, and long-dashed lines 
are gaps opened by planets of
$q=3\times10^{-5}, 1\times10^{-4}, 3\times10^{-4},$ and $1\times10^{-3}$,
respectively.  
The black lines are from \citet{bate}, and magenta lines are 
Gaussian fits, following Eq.~(\ref{gapdenprof}).  The gap opened by the 
$q=1\times10^{-3}$ planet is not well fit by a 
a Gaussian.  
}
\end{figure}

The depth and width of the gap compared to planet mass are determined 
from the results of \citet{bate}.  
In \figref{bategapfit}, we reproduce the surface density 
profiles of gap opening by planets from Figure 2 of 
\citet{bate}, having used 
DataThief\footnote{B.~Tummers, http://datathief.org}
to acquire the values.  
The unperturbed density profile is plotted as a solid line, 
and goes as $\Sigma_0(r) \propto r^{-1/2}$.
The planet is located at $r/a=1$.    
The gaps opened by planets of 0.03, 0.1, 0.3 and 1 
$M_J$ are plotted as dashed, dotted, dot-dashed, and long-dashed 
black lines, respectively.  
We fit all but the 1 $M_J$ planet to Equation (\ref{gapdenprof}), 
and plot these fits as magenta lines in \figref{bategapfit}. 
The best fit parameters are tabulated in Table \ref{fittable}.
The maximum deviation between the empirical gap profile 
and the fit, expressed as a precentage of the unperturbed disk 
density, is 0.04\% for the 0.03 $M_J$ planet, 
3\% for the 0.1 $M_J$ planet, and 
9\% for the 0.3 $M_J$ planet.  
The deviation for the 0.3 $M_J$ planet is greatest at $r/a<1$, 
indicating asymmetry of the gap that is not accounted for 
in the model presented here.  Nevertheless, fitting the gap 
to a Gaussian is a useful model for parametrizing the disk 
response to a planet without the need for running a full 
hydrodynamic simulation. 

\begin{table}[b]
\caption{\label{fittable}Best-fit parameters for 
gaps opened by planets.}
\begin{tabular}{ccccccc}
$q$\tablenotemark{1} & 
$d$ & 
$w/a$ &  
\parbox{6ex}{\centering
max.\\error\tablenotemark{2}} &
$G$ &
derived $q$ &
\parbox{12ex}{\centering
derived planet mass\tablenotemark{3}}
\\
\hline 
$3\times10^{-5}$ & 0.014 & 0.078 & 0.04\% 
& 18 & $6.7\times10^{-5}$ & 22 $M_{\earth}$
\\
$1\times10^{-4}$ & 0.56  & 0.11  & 3\% 
& 6.2 & $2.2\times10^{-4}$ & 72 $M_{\earth}$
\\
$3\times10^{-4}$ & 0.84  & 0.17  & 9\% 
& 2.5 & $6.2\times10^{-4}$ & 210 $M_{\earth}$
\\
$1\times10^{-3}$ & 
---\tablenotemark{4} & ---\tablenotemark{4} & ---\tablenotemark{4}
& 1.0 & $1.9\times10^{-3} $ & 620 $M_{\earth}$
\end{tabular}
\tablenotetext{1}{As simulated in \citet{bate}}
\tablenotetext{2}{Error$=(\Sigma-\Sigma\sub{fit})/\Sigma_0$}
\tablenotetext{3}{Actual masses used for this work.  In the text,
the masses have been rounded to 20, 70, and 200 $M_{\oplus}$
for convenience.}
\tablenotetext{4}{The gap opened by the $q=10^{-3}$ planet 
in the \citet{bate} simulation is not well-modeled by a Gaussian.}
\end{table}

In \citep{bate}, ${\cal R}=10^5$ and $H/a = 0.05$.  
Using Eq.~(\ref{eq:viscgapcrit}) as the gap opening criterion, 
this gives a gap-opening threshold of $q\sub{crit}=1.06\times10^{-3}$, 
or slightly more than 1 $M_J$.
For comparison, $r\sub{Hill}=H$ when $q=3.75\times10^{-4}$, 
so the viscous gap opening criterion gives a mass more than 
twice as large as for an inviscid disk.  
In the disk model adopted in this paper, 
at 10 AU the midplane temperature is 50K, 
$H/r=0.048$,  and $\alpha\sub{ss}=0.01$, 
so ${\cal R} = 4.4\times10^4$ and the 
gap-opening threshold is 
$q\sub{crit}=1.97\times10^{-3}$, or almost 2 $M_J$.

Given the difference in disk properties, we cannot assume that the same 
mass planets open the same size gaps in each disk.  
We estimate the masses of planet according to the following procedure.  
Defining 
\begin{equation}\label{eq:Gparam}
G\equiv \frac{3}{4}\frac{H}{r\sub{Hill}} + \frac{50}{q {\cal R}}, 
\end{equation}
we assume that $G$ is the relevant scale for determining the 
depth of the gap opened by a planet, so that   
planets with the same value of $G$ open 
similarly sized gaps.  
Then, the 0.03, 0.1, and 0.3 $M_J$ planets in the disk 
modeled by \citet{bate} have mass ratios of 
$q=3\times10^{-5}, 1\times10^{-4}$ and $3\times10^{-4}$, 
respectively, and have  $G$ values as tabulated in 
Table \ref{fittable}.  For the disk parameters adopted in this 
paper, the equivalent planet masses are then 
approximately 20, 70 and 200 $M_{\oplus}$, 
assuming that $M_{\earth}/M_{\sun}=3\times10^{-6}$.  

\begin{figure}[htbp]
\includegraphics[width=\columnwidth]{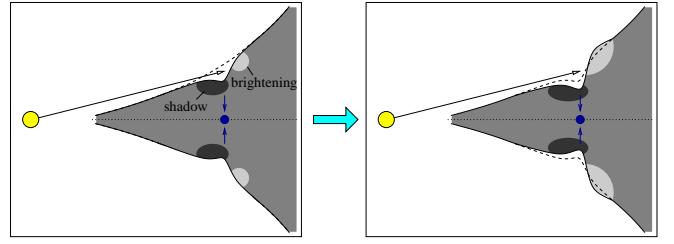}
\caption{\label{gapshadow}
Cartoon diagram of radiative feedback on disk structure.  
The star is represented as a yellow disk, and the planet by a blue dot.  
The disk surface represents a contour of constant density.
The left image shows the initial gap opened in the disk, 
with the dotted line showing the original, unperturbed disk 
surface.  Stellar illumination on the surface of the gap 
creates shadowed and brightened regions.  
Shadowing and cooling occurs in the disk trough, and the far 
side of the gap is illuminated and heated.  
The right image shows the response of the gaseous disk material to 
the cooling and heating: the shadowed region contracts and deepens the 
gap, while the illuminated far side expands and is elevated. 
}
\end{figure}

Holding $\Sigma(r)$ constant and assuming axisymmetry, 
we recalculate the temperature structure of the disk, 
using the same iterative radiative transfer method as 
described in \S\ref{initcond}.  
The effect of the gap is to create a shadow within the trough 
in density created by the gap.  The far side of the 
gap, which is now exposed to more direct stellar illumination, 
is brightened.  A schematic of this is shown in \figref{gapshadow}.  
This results in cooling within the trough and 
heating on the far wall.  The gas that composes the bulk of the disk 
material responds to this cooling and heating by contracting and 
expanding.  This changes the vertical density profile of the gap 
region, and the illuminated surface now must be recalculated.  
For this reason, the heating and cooling of the disk must be 
calculated self-consistently with the density structure of the disk.

\subsection{Observables}

Calculation of synthetic images of disks with gaps is done as in 
JC09, with the addition of including multiple scattering.
In our comparisons to Monte Carlo radiative transfer codes, we 
find that multiple scattering turns out to be an important 
component of disk emission.  
Multiple scattering generally has the property of increasing the 
brightness of a diffuse disk.  In the case of high albedos, 
the additional brightening can be 
significant.  Here, we present an estimate 
for the brightness due to secondary scattering of both 
scattered stellar irradiation and thermal emission from the disk.  
We assume isotropic scattering in these derivations.  

\subsubsection{Multiple Scattering of Stellar Irradiation}

\begin{figure}[htbp]
\plotone{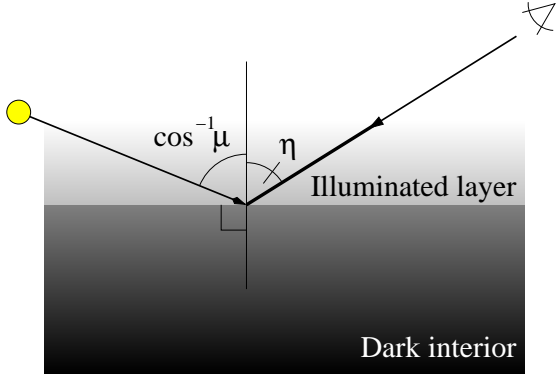}
\caption{\label{schematic}Illustration of how inclination affects 
brightness of scattered light.  
The path length through the illuminated layer represents the relative 
brightness seen by each observer. 
The cosine of the angle of incidence of stellar light at the surface 
is $\mu$.  The angle between the surface normal and the observer is $\eta$.  
}
\end{figure}

Here we consider pure scattering of stellar photons.  
From JC09, the intensity of singly scattered light of frequency $\nu$
from the disk surface is 
\begin{equation}\label{eq:sglscat}
I_1^s(\nu)=\frac{\omega_{\nu}\mu\/R_*^2\/B_{\nu}(T_*)}{4r^2(\mu+\cos \eta)},
\end{equation}
where 
$\omega_{\nu}$ is the albedo, 
$B_{\nu}(T_*)$ is the intensity emitted at the stellar surface
(here assumed to be a blackbody), 
$r$ is the distance from the star,
$\mu$ is the cosine of the angle of incidence of stellar light, 
and 
$\eta$ is the angle between the line of sight to 
the observer and normal to the surface.
The angle $\eta$ is distinct from the inclination angle 
of the overall disk, which we represent as $i$.  
The angles are illustrated in \figref{schematic}.

For multiply scattered photons, 
we first need to determine the diffuse scattered 
radiation field within the disk, $J_s$.  
This was solved for in \citet{paper1,paper2} 
using plane-parallel equations for radiative transfer as 
\begin{eqnarray}
J_s &=& \frac{B_{\nu}(T_*) R_*^2}{4 l^2}
  \frac{\mu \omega_{\nu}}{1-3g^2\mu^2}
\nonumber \\ & & \quad \times
    \left[ \frac{2+3\mu}{1+2g/3}  \exp(-g\tau_{\perp})
      - 3\mu \exp(-\tau_{\perp}/\mu) \right]
\end{eqnarray}
where $g=\sqrt{3(1-\omega_{\nu})}$
(see also \citet{vertstruct}).

Given $J_s$, then the brightness of photons scattered to the observer 
after two or more scatterings is 
\begin{equation}
I_2^s(\nu) = \int_0^{\infty} \omega_{\nu} \exp(-\tau\sub{obs}) J_s \, d\tau\sub{obs}
\end{equation}
where $\tau\sub{obs}$ is the line-of-sight optical depth.  
If $\tau_{\perp}$ is the optical depth perpendicularly 
below the surface of the disk, and $\eta$ is the 
angle of the observer with respect to the disk normal, 
then $\tau\sub{obs}=\tau_{\perp}/\cos\eta$.  
Then, 
\begin{eqnarray}
I_2^s &=&   
\frac{B_{\nu}(T_*) R_*^2}{4 l^2}
  \frac{\mu \omega^2}{1-g^2\mu^2} 
\times \nonumber \\ & & \quad
    \left[ \frac{2+3\mu}{(1+2g/3)}
    \frac{1}{(1+g\cos\eta)}
      - \frac{3\mu}{(1+\cos\eta/\mu)}
\right]
\end{eqnarray}
and
\begin{eqnarray}
I_{\nu}^s &=& I_1^s + I_2^s = 
\frac{\omega_{\nu}\mu\/R_*^2\/B_{\nu}(T_*)}{4r^2(\mu+\cos \eta)}
\times \nonumber \\ & & \quad 
\left\{1 + 
\frac{\omega}{1-g^2\mu^2} 
  \left[\frac{(2+3\mu)(\mu+\cos\eta)}{(1+2g/3)(1+g\cos\eta)} - 3\mu^2
\right]
\right\}.\label{eq:multscat}
\end{eqnarray}

\subsubsection{Scattered Thermal Emission}


Assuming a plane-parallel disk atmosphere, 
the radiative transfer equation is 
\begin{equation}\label{intensity}
\cos\eta \frac{\partial I_{\nu}}{\partial\tau_{\nu}} = I_{\nu}-S_{\nu}
\end{equation}
where $\eta$ is the viewing angle with respect 
to the line perpendicular to the disk, and 
$\tau_{\nu}$ is the total extinction perpendicular to the surface, 
including both absorption and scattering.  
For scattering and absorption, the source function is 
\begin{equation}\label{source}
S_{\nu} = \omega_{\nu} J_{\nu} + (1-\omega_{\nu}) B_{\nu}
\end{equation}
where $\omega_{\nu}$ is the wavelength-dependent albedo and 
$B_{\nu}=B_{\nu}(T)$ is the local thermal emission.  
Adopting the Eddington approximation, 
\begin{equation}\label{diffeq}
\frac{1}{3} \frac{\partial^2 J_{\nu}}{\partial\tau_{\nu}^2} = 
(1-\omega_{\nu})(J_{\nu}-B_{\nu}).
\end{equation}

We have $B_{\nu}$ from the temperature structure of the disk, and 
we can integrate Eq.~\ref{diffeq} to get $J_{\nu}$.  
Finally, we integrate Eq.~\ref{intensity} to get the emitted 
intensity at the surface of the disk.  

In the case of an isothermal slab of finite optical thickness
($\tau\sub{max}$), we can 
set the boundary conditions at $\tau=0$ and $\tau=\tau\sub{max}$ 
\begin{eqnarray}
J_{\nu}(\tau=0) 
&=& \frac{1}{\sqrt{3}}\frac{\partial J}{\partial \tau}(\tau=0) \\
J_{\nu}(\tau=\tau\sub{max}) 
&=& -\frac{1}{\sqrt{3}}\frac{\partial J}{\partial \tau}(\tau=\tau\sub{max})
\end{eqnarray}
\citep{rybickilightman}.  Solving, 
\begin{eqnarray}
\label{Jisotherm}
J_{\nu} &=& B_{\nu}(T) \left\{
1 
\right. \\ & & \left.
- \frac{\exp\left[(\tau-\tau\sub{max})\sqrt{3(1-\omega_{\nu})}\right]
+\exp\left[-\tau\sqrt{3(1-\omega_{\nu})}\right]
}{
\left(1-\sqrt{1-\omega_{\nu}}\right)
\exp\left[-\tau\sub{max}\sqrt{3(1-\omega_{\nu})}\right]
+1+\sqrt{(1-\omega_{\nu})}
}
\right\}. 
\nonumber
\end{eqnarray}
If we assume that $T$ varies slowly with $\tau$, then we can 
approximate $J_{\nu}$ by Eq.~(\ref{Jisotherm}) locally, 
as done in \citet{dalessio3}
Then the emergent intensity can be found by integrating 
Eq.~(\ref{intensity}), assuming that the background intensity is $0$.  
To simplify the calculation, we assume that the disk 
is locally plane parallel, so $\eta=i$, the inclination angle.  
For a face on disk, $\eta=0$.

The total intensity is then $I_{\nu} = I_{\nu}^s + I_{\nu}^t$, 
although as a general rule $I_{\nu}^2$ 
dominates in optical to mid-IR, and $I_{\nu}^t$ 
dominates at longer wavelengths.  

\section{Validation}

The JC radiative transfer model for calculating the 
disk structure for this work is 
fully three-dimensional, in the sense that radiation impinging 
on the surface of the disk is allowed to propagate in all directions 
throughout the disk.  However, it relies on a locally one-dimensional analytic 
solution to the radiative transfer equation and adopts 
a number of simplifying assumptions such as mean opacities 
\citep[see also][]{paper1,paper2}.  
The novel approach presented in this paper is the iterative calculation
of the self-consistent density and temperature structure.  
As shown by the significant temperature perturbations created 
by shadowing and illumination on gaps, the self-consistency 
is an important aspect to consider in the analysis of radiative
transfer in disks.  

In order to validate the radiative transfer prescription adopted 
in this paper, we compare our results to a Monte Carlo 
radiative transfer calculation on the final disk density structure
\citep{2011arXiv1110.4166T}.  
The Monte Carlo approach is similar to that of \citet{2006Pinte_etal}.
We find the radiative equilibrium temperatures by
following a large number of photon packets from the star through
scattering, absorption and re-emission until escape to infinity.  In
this way the energy is conserved exactly.  We sum the radiation energy
absorbed all along the packet paths \citep{1999Lucy} and relax
to equilibrium by choosing each re-emitted packet's frequency to
adjust the local radiation field for the updated temperature 
\citep{2001BjorkmanWood}.
The scattering is assumed to be isotropic.
Each of the calculations shown involves $10^9$ photon packets.

In the following section, as we discuss our results for the 
disk temperature structure and observable quantities, we 
compare the JC model 
to the results found by the Monte Carlo (MC) method.

\section{Results}

\subsection{Gap Structure}\label{gapstruct}

\begin{figure}[tbh]
\plotone{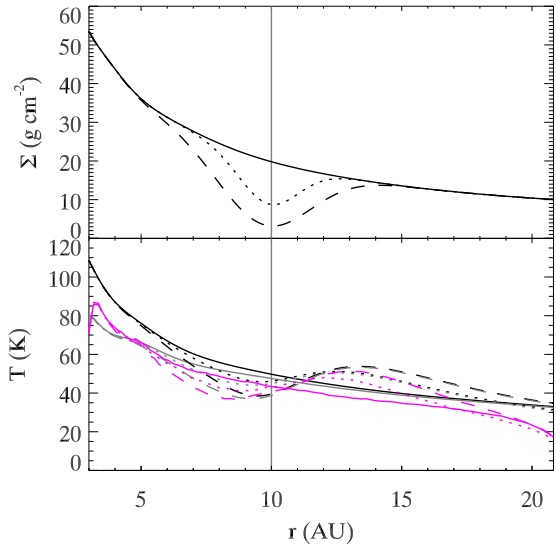}
\caption{\label{dentemp}
Surface density (top) and midplane temperature (bottom) profiles 
for a disk with and without a gap created by a planet at 10 AU\@. 
Solid, dotted, and dashed lines indicate 
planet masses of 0, 70, and 200 $M_{\oplus}$, respectively.  
The thermal perturbation is caused by shadowing and illumination
by stellar irradiation at the disk surface.  
The black lines indicate the results of the JC models 
while the gray lines show JC models excluding viscous heating 
at the midplane.  
The magenta lines on the temperature plot show the results of 
a Monte Carlo radiative transfer calculation on the same 
three-dimensional density structure.
}
\end{figure}

The surface density profiles imposed on the disk for 
0, 70, and 200 $M_{\earth}$ planets are shown in the upper 
plot of \figref{dentemp}, as solid, dotted, and dashed lines,
respectively.  
The calculated midplane temperatures for the corresponding disk 
models are summarized in the lower plot of 
\figref{dentemp}, in black for the JC models, and 
in magenta for the MC calculations. 

The Monte Carlo calculations include only the disk annulus between
cylindrical radii 3~and 20.8~AU\@.  For computational speed, the highly
optically-thick material within 3~AU is omitted.  Without the
obscuring material, the wall facing the star at 3~AU would become
unexpectedly hot.  We mitigate this by immediately discarding any
Monte Carlo photon packet absorbed or scattered for the first time in
the innermost column of grid cells.  Some residual excess heating
comes from the fact that omitting the inner disk reduces the column to
the star even in the atmosphere, where the innermost cells are
optically-thin.  This residual excess is significant only inside about
4~AU\@.  On the other hand, truncating the disk model at the outer
radius reduces temperatures, since photon packets reaching 20.8~AU can
escape to infinity, radiatively cooling the midplane.  Considering
these approximations together, we believe the Monte Carlo results
between 4~and 18~AU accurately represent the solution that would be
obtained at greater computational expense in a disk model continuing
inward to the sublimation radius and outward to the disk's far edge.

The JC temperature is consistently higher than the MC temperatures
overall.  
One cause of this discrepancy is that the JC model includes viscous 
heating at the midplane, so that the final temperature is 
\begin{equation}
T = (T_r^4 + T_v^4)^{1/4}
\end{equation}
where $T_r$ and $T_v$ are the temperatures 
resulting from stellar irradiation and viscous heating, respectively.
The viscous temperature is given by 
\begin{equation}
T_v^4 = \frac{9}{32\pi}
\frac{GM_{\star}\dot{M}}{\sigma_B r^3}
	\left[1-\left(\frac{R_{\star}}{r}\right)^{1/2}\right]
(\tau_d+2/3)
\end{equation}
where $\sigma_B$ is the Stefan-Boltzmann constant and $\tau_d$ 
is the vertically integrated optical depth, 
$\tau_d=-\int_{\infty}^{z}\chi_R\rho\, dz$,
with respect to
$\chi_R$, the Rosseland mean opacity.  

In \figref{dentemp}, we plot $T_r$ in grey to show the effect of 
removing viscous heating.  Viscous heating contributes more 
at smaller $r$, and is neglible beyond $\sim$15 AU\@.   
Interior to 5 AU, $T_r$ is less than that calculated by the MC 
model, and outside of 5 AU, the MC temperatures fall more 
rapidly than the JC temperatures.  Thus, viscous heating 
can explain the discrepant temperatures only at small radii.  
Another possible explanation is that the difference lies in the 
treatment of opacities in the two models, as will be described below.  

The gap opened by the 20 M$_{\oplus}$ planet is less than 2\% in 
depth, and produces no more than 2\% excursions in temperature 
from the unperturbed disk, so small as to be negligible.  
We thus set 20 M$_{\oplus}$ as the lower bound on a 
detectable planet in this disk at 10 AU, or more generally, 
planets with 
$q \leq 0.036\,q\sub{crit}$ or $G \geq 18$ 
are undetectable, 
whereas planets with 
$q \geq 0.12\,q\sub{crit}$ or $G \leq 6$ 
do significantly perturb the disk.  

In contrast to the results of JC08, the midplane temperature 
is significantly affected by gap opening.  
JC08 considered planets up to 50 $M_{\oplus}$ in the 
absence of a gap, but here we find that a 70 $M_{\oplus}$ planet should  
open a significant gap.  In a disk with lower viscosity, 
a 50 $M_{\oplus}$ might open a similarly sized gap.  
This shows the importance of including large scale non-linear 
dynamical interactions in planet-disk models.  

\figref{dentemp} illustrates that a gap in a disk creates an
S-shaped perturbation to the midplane temperature profile.  
The temperature at the position 
of the planet is lowered, although the temperature 
minimum is inward of the planet position.  
This is because cooling at the gap minimum is mitigated 
by heating on the far gap wall.  
In the JC model, 
the minimum and maximum temperature deviations at the midplane 
for the 70 $M_{\oplus}$ planet 
are $-6$ K ($-11\%$) at 8.9 AU and $+8$ K ($+19\%$) at 13.4 AU, respectively.  
In the MC model, these values 
are $-6$ K ($-13\%$) at 8.0 AU and $+9$ K ($+25\%$) at 13.1 AU, 
respectively.  
For the 200 $M_{\earth}$ planet, the values are 
$-13$ K ($-25\%$) at 9.1 AU and $+12$ K ($+29\%$) at 13.9 AU 
in the JC model, and 
$-11$ K ($-23\%$) at 7.9 AU and $+14$ K ($+38\%$) at 13.4 AU\@. 
In general, the magnitude of heating and cooling in Kelvins is 
comparable, but the percentages differ because the unperturbed 
temperature of the MC model is smaller.  The positions 
of the minima and maxima in the JC model are slightly inward of the 
minima and maxima in the MC model, which might be caused 
by the same reasons that give rise to the lower midplane 
temperatures in the MC model.  

\begin{figure}[htbp]
\begin{center}
\plotone{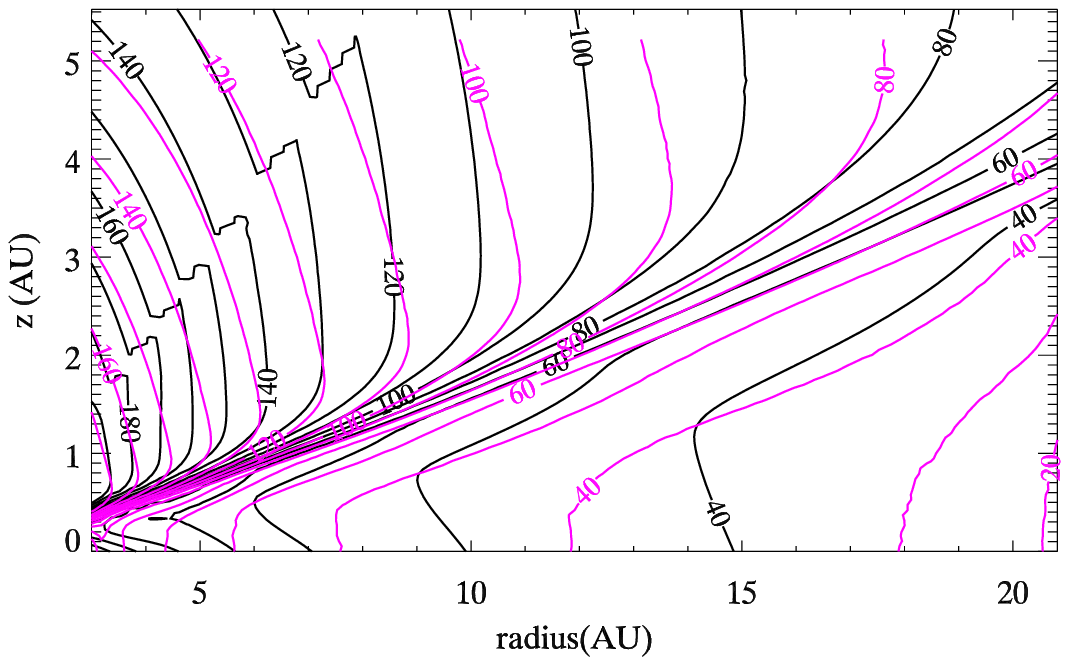}\\
\plotone{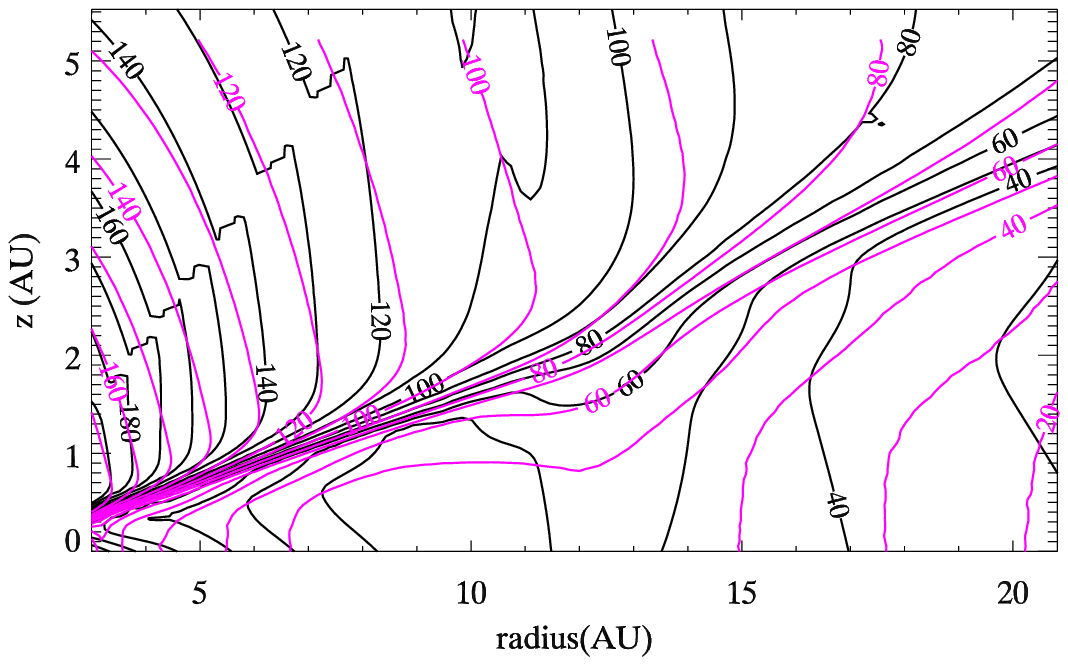}\\
\plotone{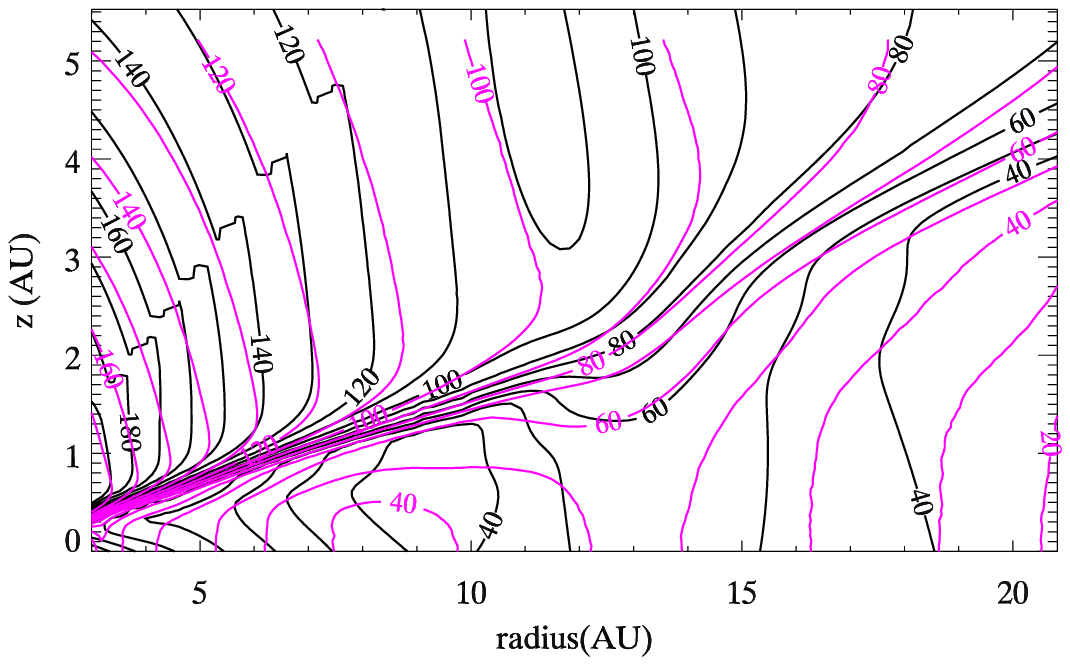}
\end{center}
\caption{\label{tempmaps}
Temperature cross-sections of disk models with and without gaps.  
Black and magenta lines show the JC and MC 
methods, respectively, of calculating radiative heating.
Top: A disk without a planet or a gap.
Middle: A disk with a gap created by a 70 $M_{\earth}$ planet.  
Bottom: A disk with a gap created by a 200 $M_{\earth}$ planet.  
}
\end{figure}

The full radial and vertical temperature structure of our models 
are shown in \figref{tempmaps}.  
The vertical axis is stretched compared to the horizontal axis 
in these plots.  The top, middle, and bottom plots show 
gaps created by 0, 70, and 200 $M_{\earth}$ planets, respectively.  
The JC models are plotted in black, while the 
MC models are plotted in magenta.  

The abrupt shifts in temperature in the upper left corners 
of the JC model are purely a numerical artifact.  
This region is extremely low density, so this region is considered 
to be outside the bounds of the simulation, and the temperature 
is assigned based on zero optical depth and a radial dependence.  
Since these regions are 
very low density, the temperature differences have no 
effect on hydrostatic equilibrium nor on the observables, and 
do not affect the remainder of the analysis presented in this paper, 
though the topic should be addressed in future studies. 

The disk surface is coincident with regions where the temperature changes 
rapidly in the vertical direction, i.e.~where the temperature 
contours lie close together.  Above the surface, the JC model
is generally warmer than the MC model.  
However, at or just below the surface, the MC model is generally 
warmer.  Closer to the midplane, the JC model again becomes warmer, 
as discussed previously.  
The warmer temperatures in the MC model below the disk surface   
indicate that heating by stellar irradiaton 
penetrates deeper in the MC 
model than the JC model.  The probable cause of this
and other differences in temperatures between models, 
is that the JC model calculates the propagation 
of stellar photons into the disk using mean opacities, which treats 
all stellar photons as if they have the same opacity.  
In the MC models, 
opacity varies with wavelength, so photons with wavelengths 
longward of the stellar peak have a lower opacity and can heat 
the disk to deeper depths.  The inclusion of viscous heating 
in the JC model leads to the increase in temperature 
at the midplane.  

When a gap is imposed on the disk, 
as in the lower two plots of \figref{tempmaps}, 
shadowing and cooling within the gap resulting in the 
temperature contour lines dipping downward.  
The depth of these dips are comparable in both the 
JC and MC models, although the exact 
shape differs.  This difference may again be explained by 
the the choice of mean opacities versus wavelength-dependent 
opacities.  Allowing longer wavelength photons to penetrate further 
into the disk would have the effect of smearing out the 
temperature differences casued by surface effects.  

A close look at the surface temperature contours 
at $r>20 AU$ reveals that the surface temperatures are lower for the 
a gapped disk in both the JC and MC models.  
This is a result of shadowing of the outermost radii by 
the puffed up outer edge of the gap.  
This shadowing and cooling effect is quite real, 
but since the shadow extends beyond the simulation 
boundaries, quantifying this effect further is beyond the scope 
of this paper.  
A similar shadowing by the illuminated outer edge of a gap created 
by Jupiter by \citet{2011arXiv1110.4166T}.

\subsection{Simulated Images}\label{sec:images}

\begin{figure}[htbp]
\includegraphics[width=\columnwidth]{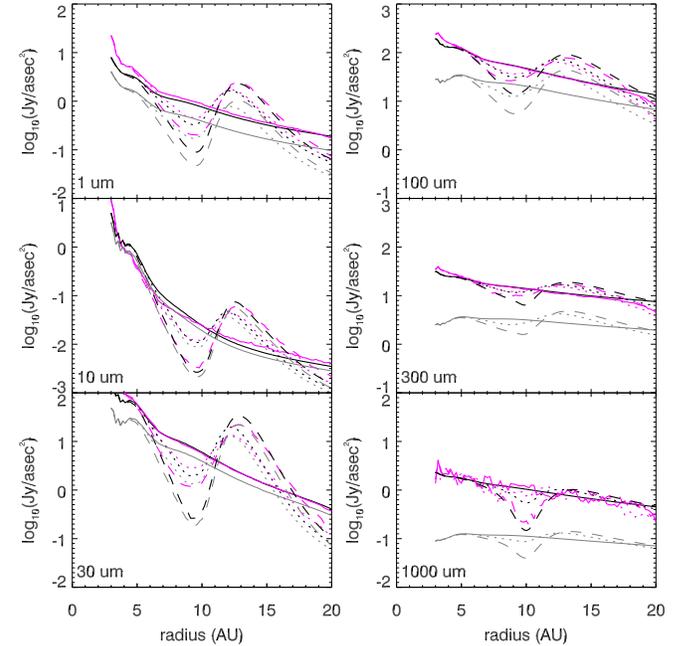}
\caption{\label{gapprofiles}
Radial surface brightness profiles of face-on disks with and without gaps 
at the wavelengths 
imaged in Figures \ref{faceonshort} \& \ref{faceonlong}, 
as indicated.  
The gapless disk surface brightness are plotted with solid lines.
Surface brightness profiles of gaps created by 
70 and 200 $M_{\oplus}$ planet are plotted as dotted and dashed 
lines, respectively.
Black lines indicate the JC models including multiple scattering, 
while gray lines show only single scattering or direct emission.  
For comparison, results from MC 
calculations on the identical disk density structure are overplotted 
as magenta lines.  The 1 mm brightness profiles is noisy because 
of low photon statistics.  
MC calculations give consistently higher brightnesses
because of multiple scattering and temperature differences, 
as discussed in the text, 
}
\end{figure}

The radial surface brightness profiles for each of the disks 
are shown in \figref{gapprofiles} at wavelengths of $1-1000$ $\mu$m.  
The JC models are plotted in black, 
and the Monte Carlo models are plotted in magenta.  
The brightness profiles of the unperturbed disk are plotted 
as solid lines, and gaps created by 70 and 200 $M_{\earth}$ planets 
are plotted as dotted and dashed lines, respectively.  
We show general good agreement between the models.  
The noise in the brightness profiles at 1000 microns in the MC models 
is caused by low photon statistics.  
This is because the 1000 microns wavelength lies out in the tail of the
thermal distribution.  Most photon packets have wavelengths near
the peak emission of star or disk.   

For comparison, we plot as gray lines in \figref{gapprofiles} 
the JC brightness profiles for single
scattering and direct thermal emission only, as calculated in JC09.
Inclusion of multiple scattering of both scattered stellar photons and
thermally emitted photons bring the JC brightness profiles in line
with the MC ones.  
At 1 micron, the albedo is $\omega_{\nu}=0.91$ and the disk emission 
is effectively purely scattered light, so the disk about
twice as bright.  At 10 microns, $\omega_{\nu}=0.45$, and scattered 
photons and thermally emitted photons are roughly equal.  
Because of the relatively low albedo, 
multiple scattering increases the brightness 
by about $30\%$ averaged over radius.  
At 30 microns, $\omega_{\nu}=0.45$ also, but 
because thermal emission is dominant at the wavelenght, 
the overall brightness is increased by about $70\%$.  
Multiple scattering is more effective at increasing thermal 
emission because it allows photons emitted in optically thick 
regions to percolate up through the disk.  Scattered light, 
on the other hand, is limited to the photons incident on the 
surface of the disk.  As the albedo increase, the amount of 
brightening increases: at 100, 300, and 1000 microns, 
the $\omega_{\nu}=0.57, 0.82,$ and $0.95$, and the brightness 
is increased by factors of $2.6$, $4.9$, and $9.8$, respectively, 
as averaged over radius.  

The perturbation from the gap shadow is evident across all wavelengths, 
reflecting the cooling and heating that take place in the shadowed 
trough and brightened rim of the gap.  The puffing up of the 
far rim of the gap is significant enough to shadow and cool the 
outer disk, as evident in the steeping of the brightness profile 
toward 20 AU at $1-100$ microns.  The brightening of the far rim of the gap 
in the 1 $\mu$m brightness profile 
is consistent with the results of 
\citet{2006Varniere_etal}, with feedback from cooling and shadowing 
resulting in a puffing up of the gap edge.  

Despite the offset in the midplane temperature maxima and minima, 
the appearance of the gap at all wavelengths is consistent.  
The JC model tends to overpredict the depth of the gap, particularly 
at 1 micron, and from $30-300$ microns.  At 1 micron, this 
maybe because photons can scatter from the back wall of the gap 
into the gap, and this is not captured in the JC model.   
From $30-300$ microns, the difference may arise from the detailed 
differences in the gap temperature structure between the JC and MC models.  
At least qualitatively the models agree.

Having validated the JC model by comparison to the MC model, 
the remainder of our analysis will focus on the JC model.  

\begin{figure*}[htbp]
\includegraphics[width=0.32\textwidth]{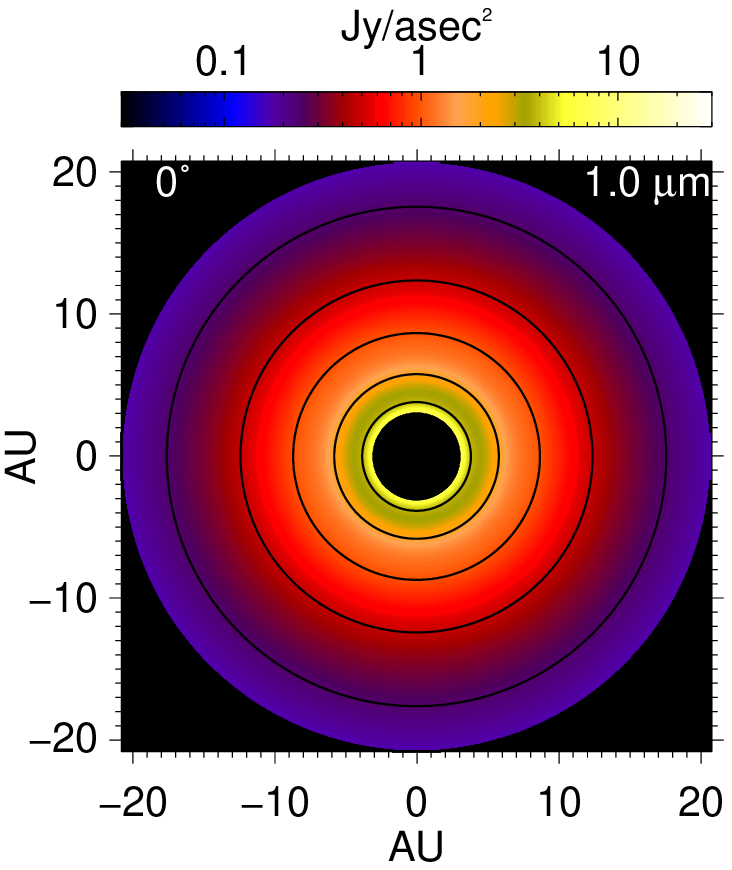}
\hfill
\includegraphics[width=0.32\textwidth]{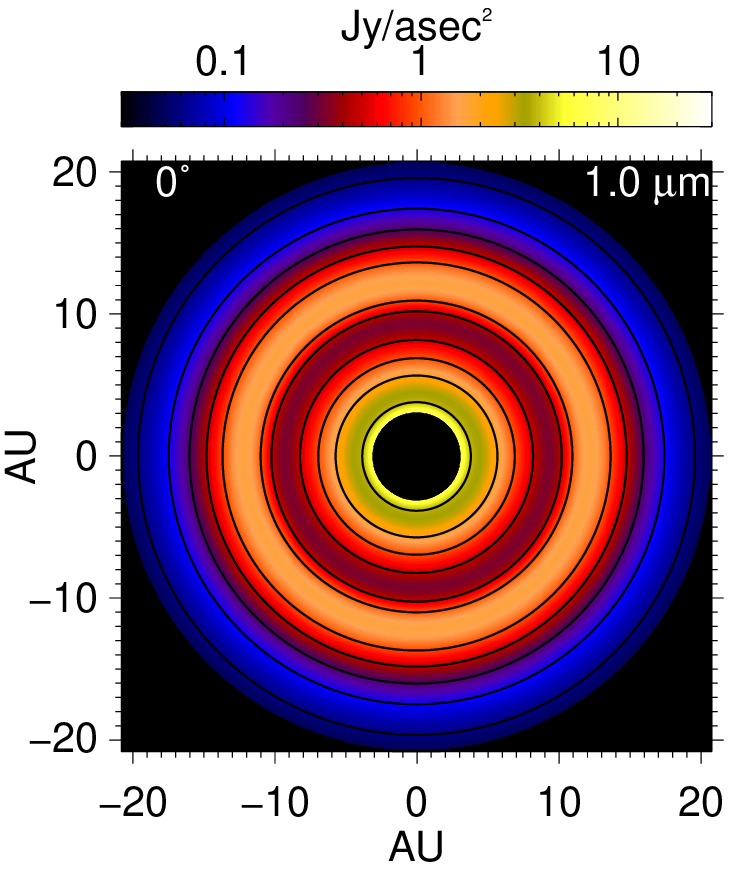}
\hfill
\includegraphics[width=0.32\textwidth]{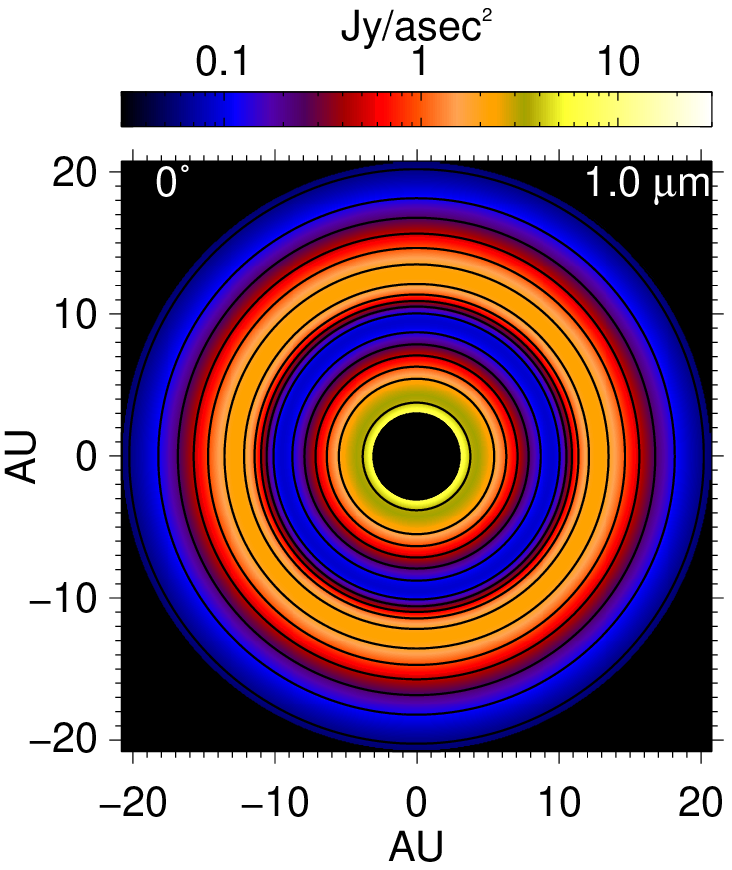}
\\
\includegraphics[width=0.32\textwidth]{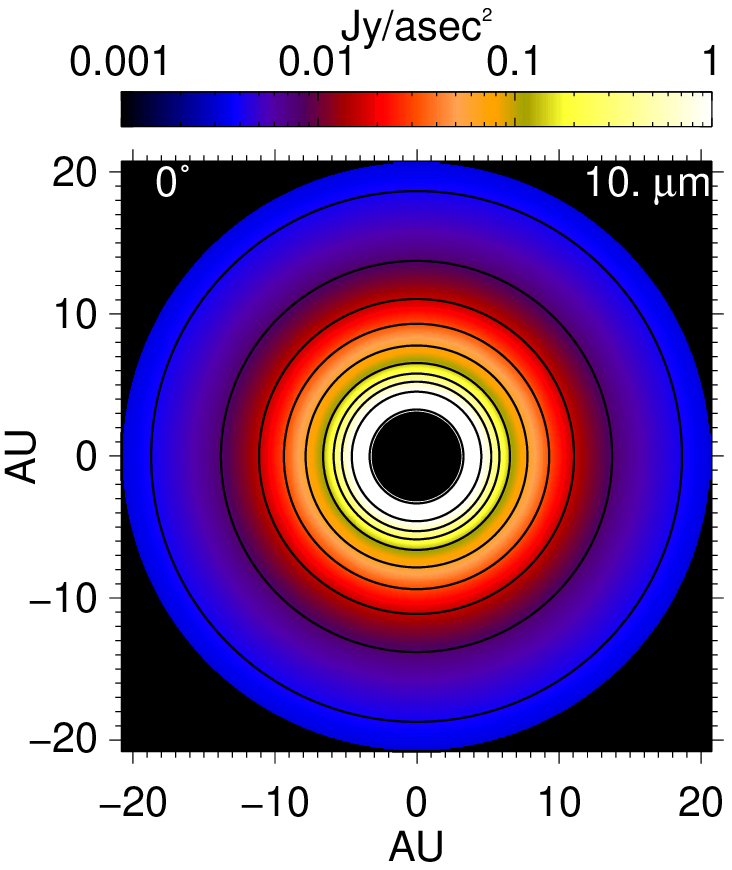}
\hfill
\includegraphics[width=0.32\textwidth]{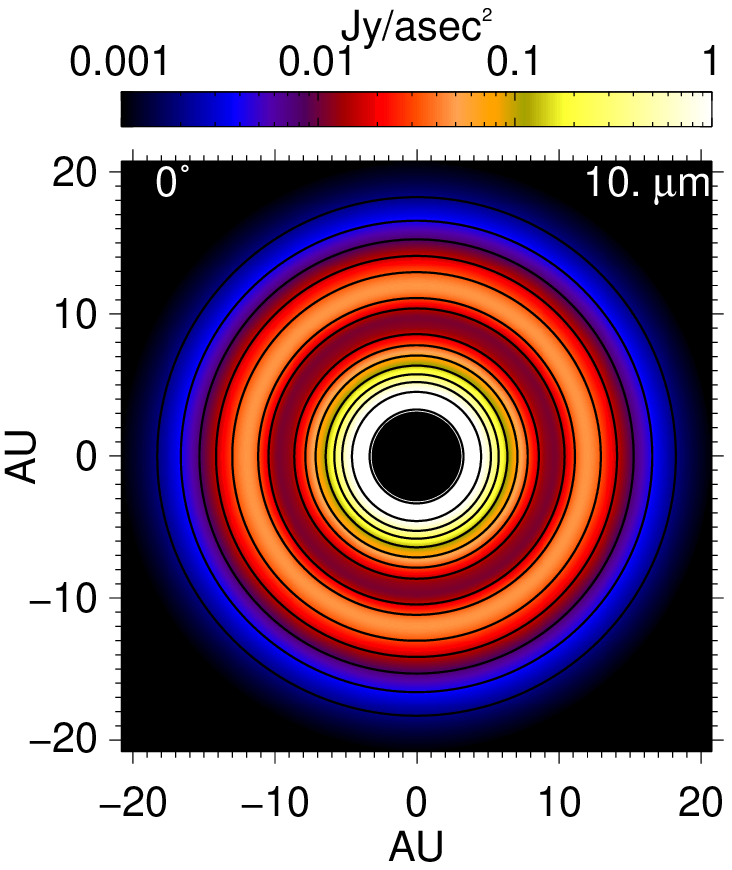}
\hfill
\includegraphics[width=0.32\textwidth]{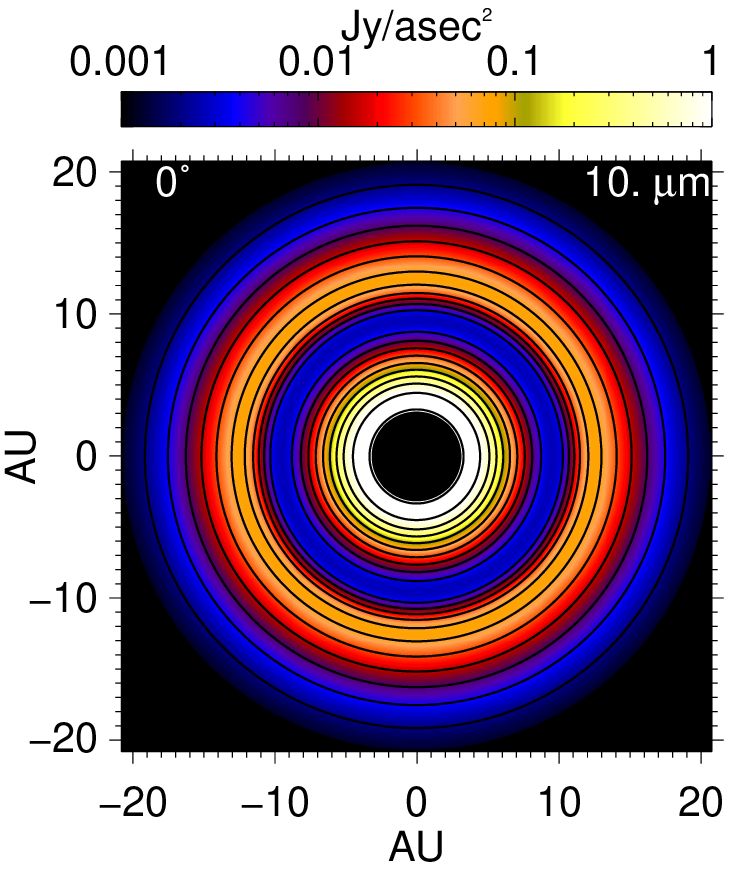}
\\
\includegraphics[width=0.32\textwidth]{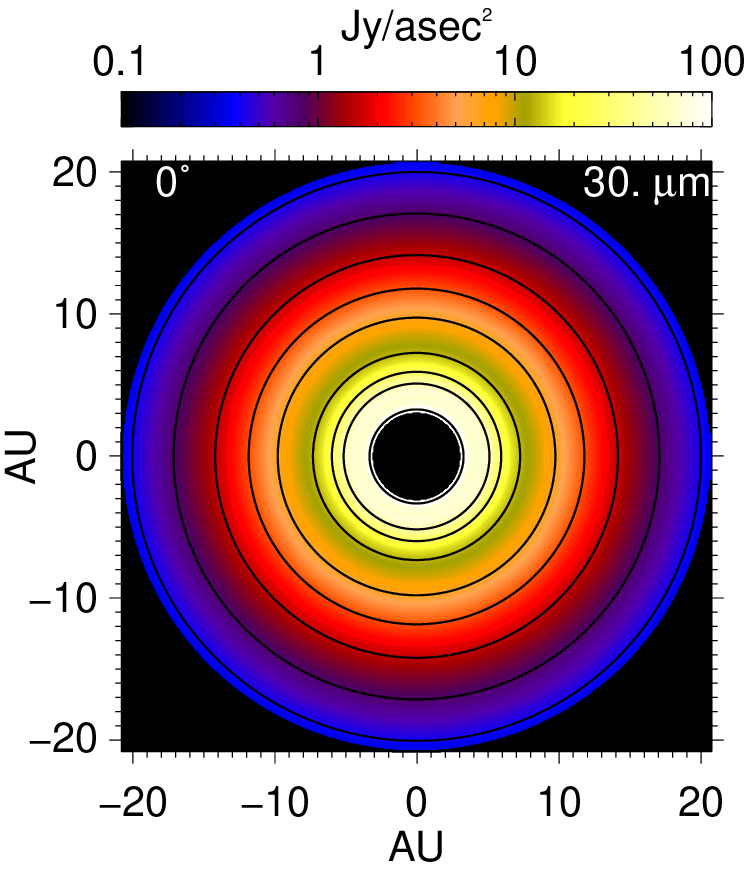}
\hfill
\includegraphics[width=0.32\textwidth]{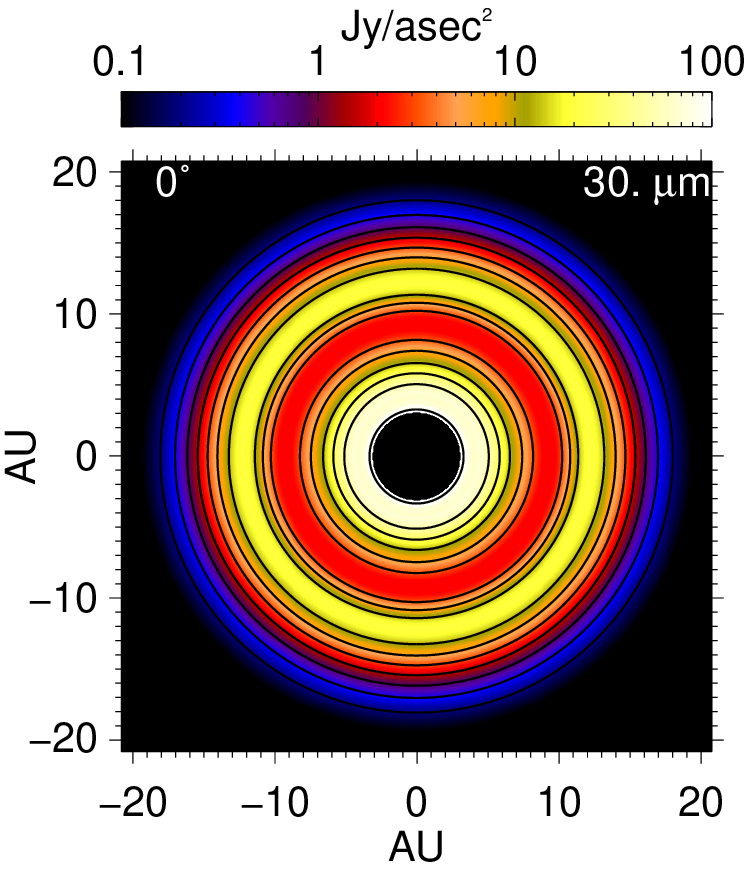}
\hfill
\includegraphics[width=0.32\textwidth]{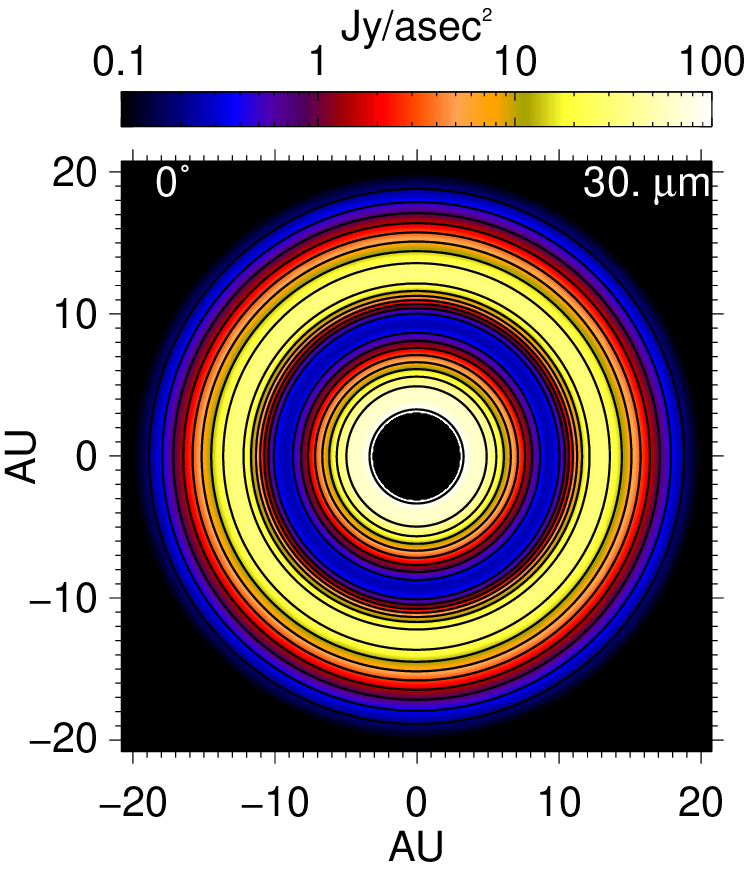}
\caption{\label{faceonshort}
Simulated images of disks from JC models with and without gaps at 
1 (top), 10 (middle), and 30 (bottom) microns.  
The left, center, and right 
images in each row show disks with 0, 70, and 200 M$_{\oplus}$ planets, 
respectively.  The contours are spaced at intervals of factors of two
in brightness, and the colors trace brightness 
according to the displayed color bars.  
}
\end{figure*}

\begin{figure*}[htbp]
\includegraphics[width=0.32\textwidth]{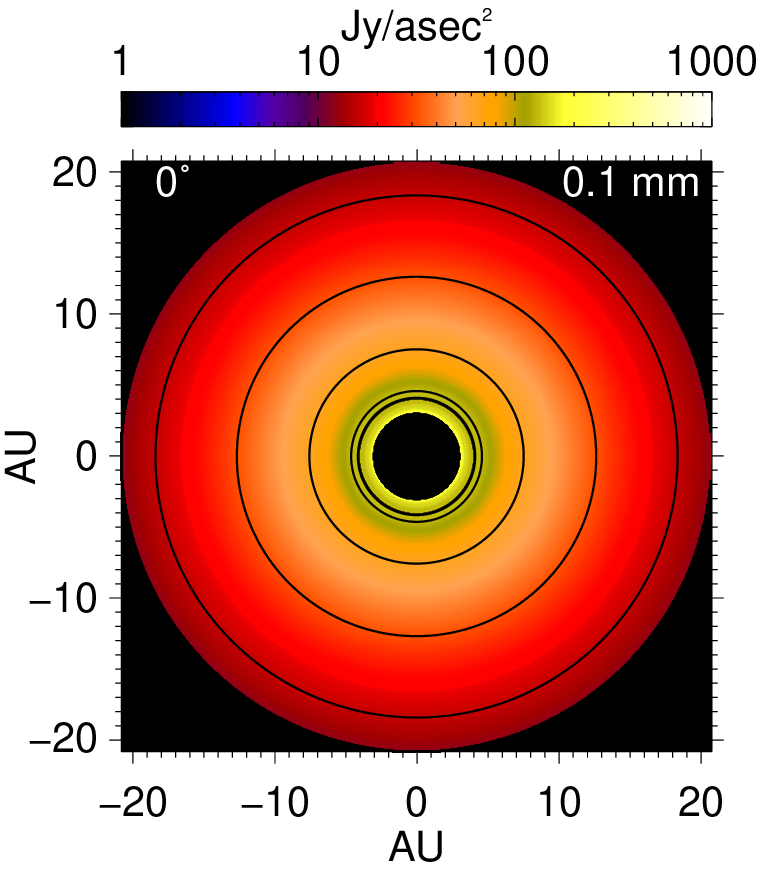}
\hfill
\includegraphics[width=0.32\textwidth]{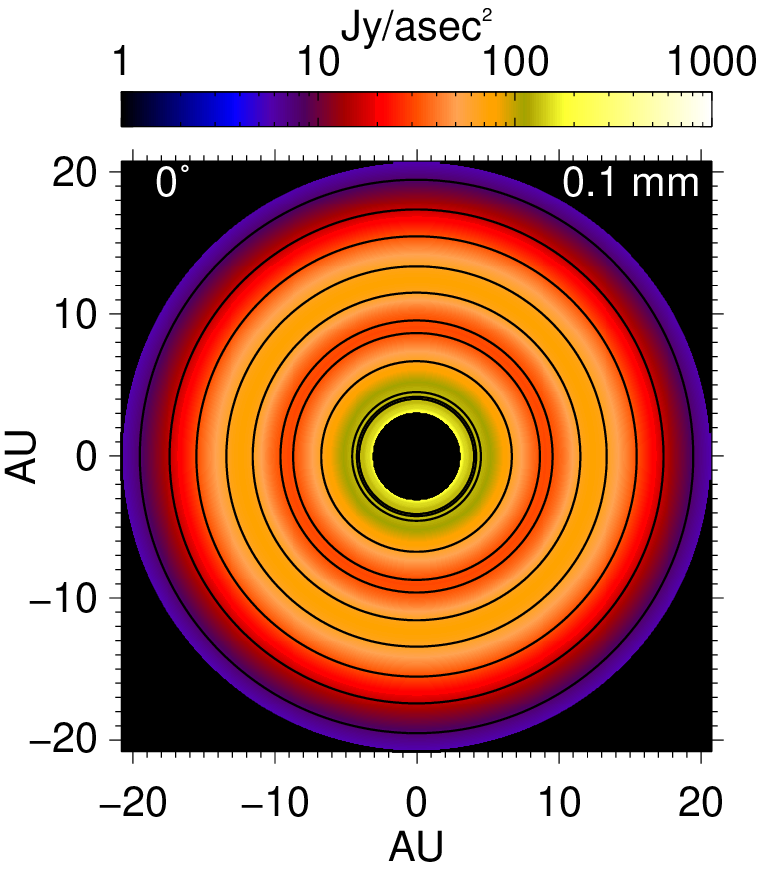}
\hfill
\includegraphics[width=0.32\textwidth]{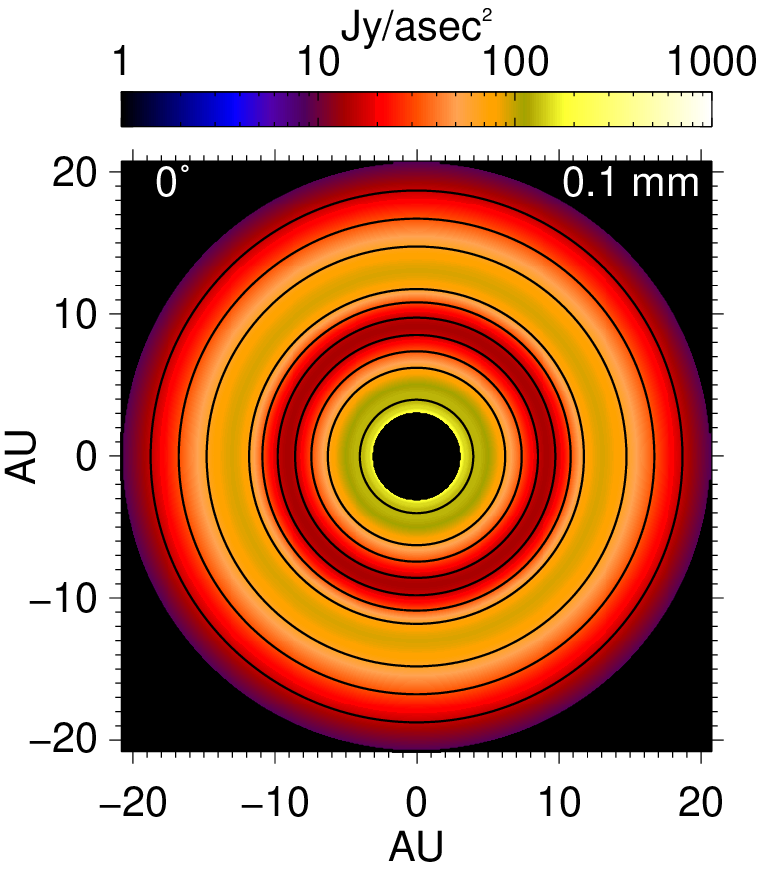}
\\
\includegraphics[width=0.32\textwidth]{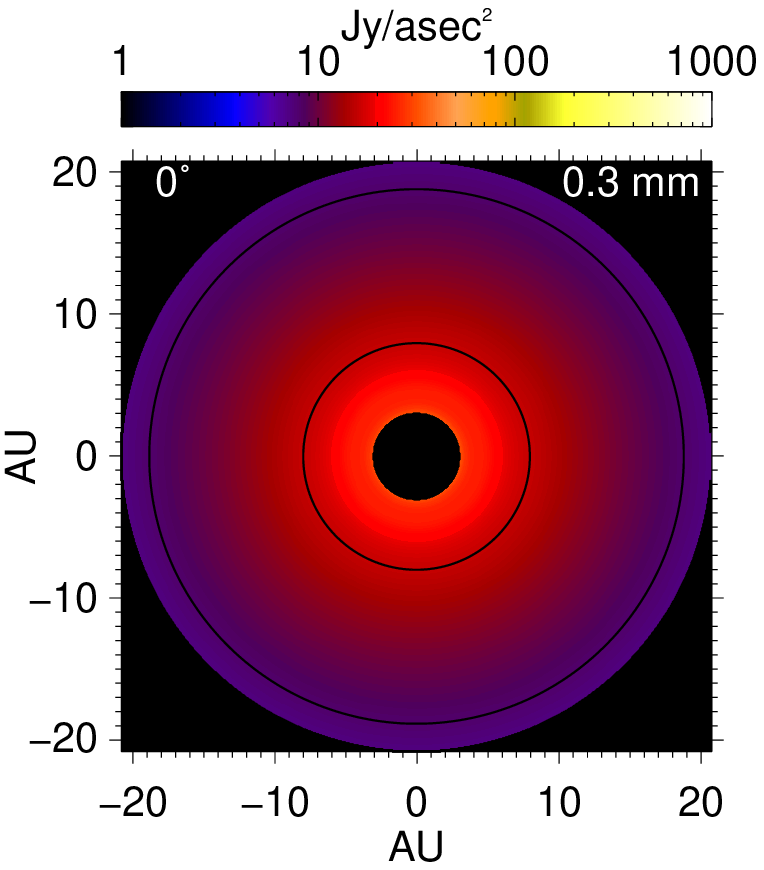}
\hfill
\includegraphics[width=0.32\textwidth]{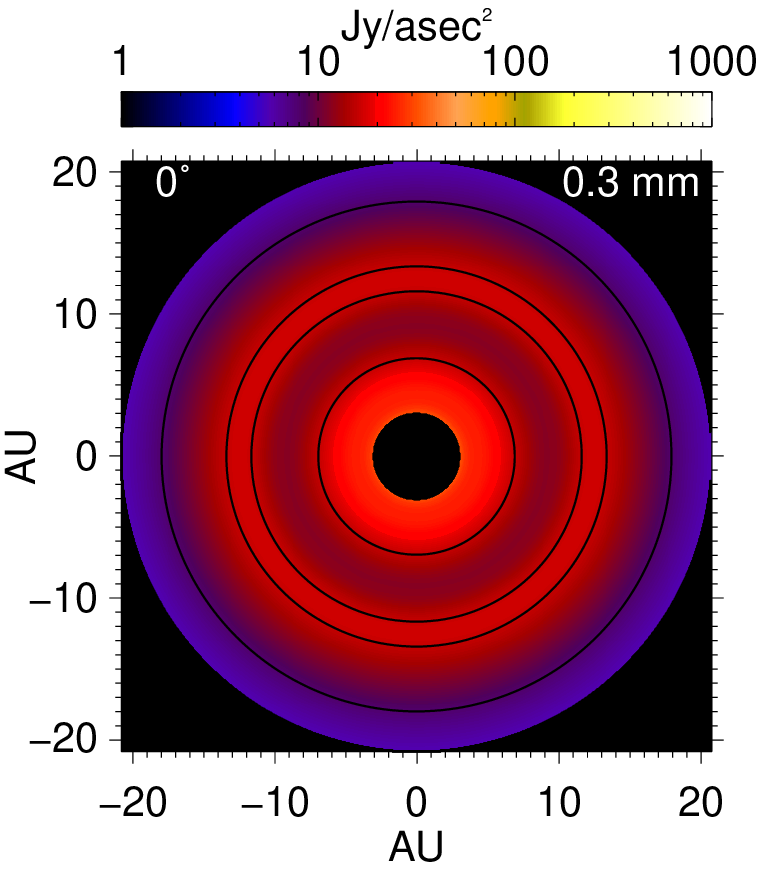}
\hfill
\includegraphics[width=0.32\textwidth]{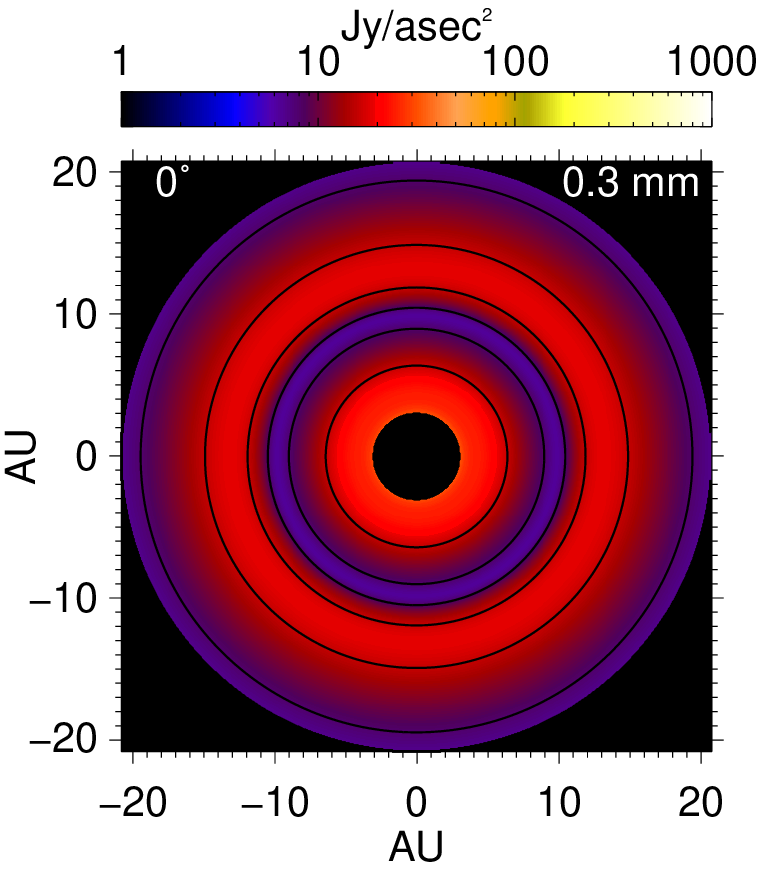}
\\
\includegraphics[width=0.32\textwidth]{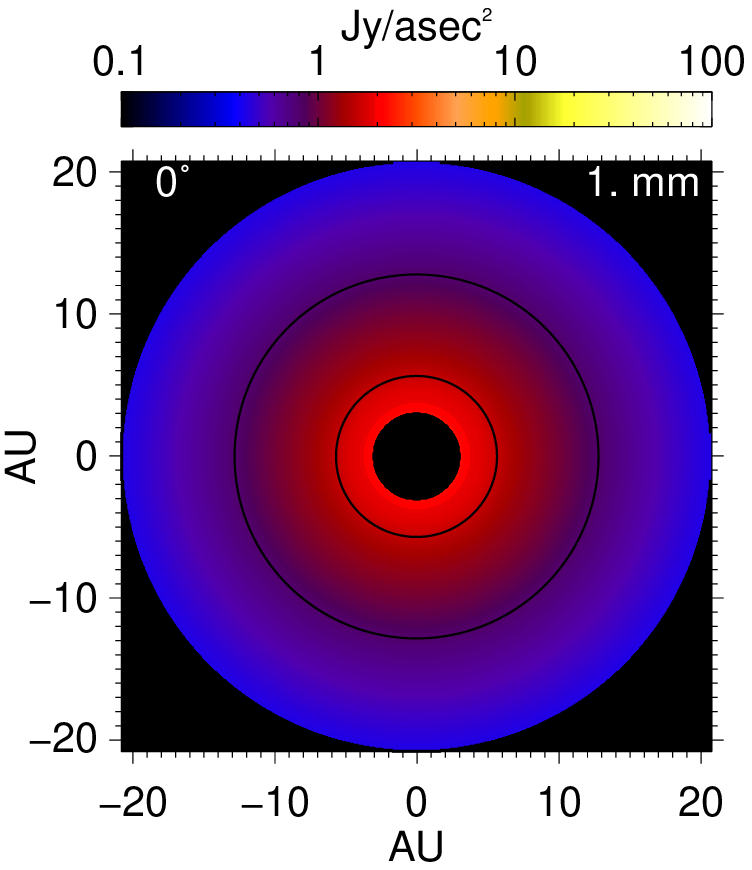}
\hfill
\includegraphics[width=0.32\textwidth]{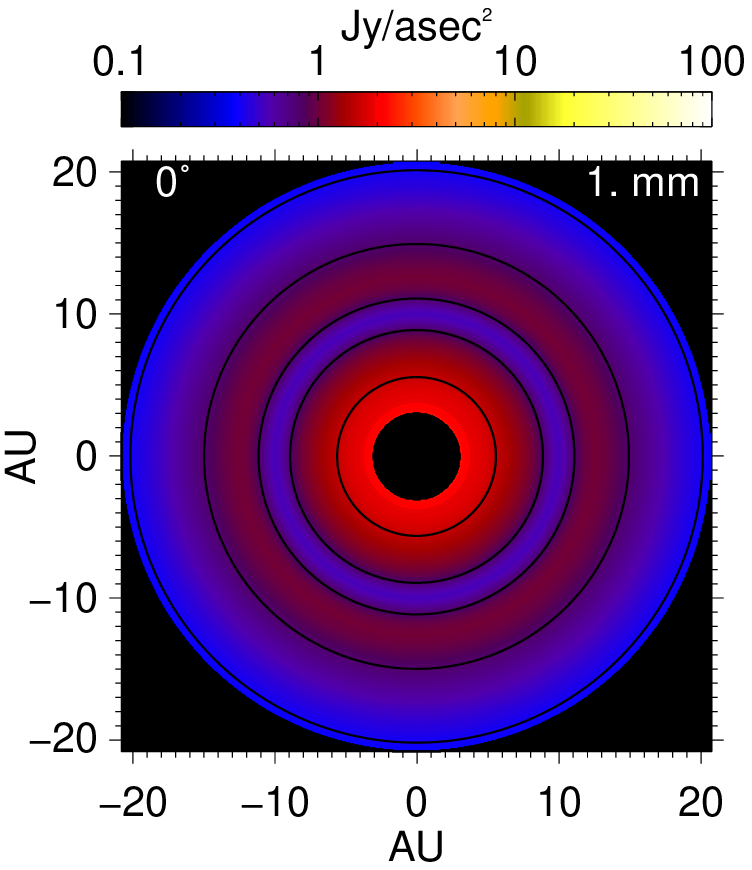}
\hfill
\includegraphics[width=0.32\textwidth]{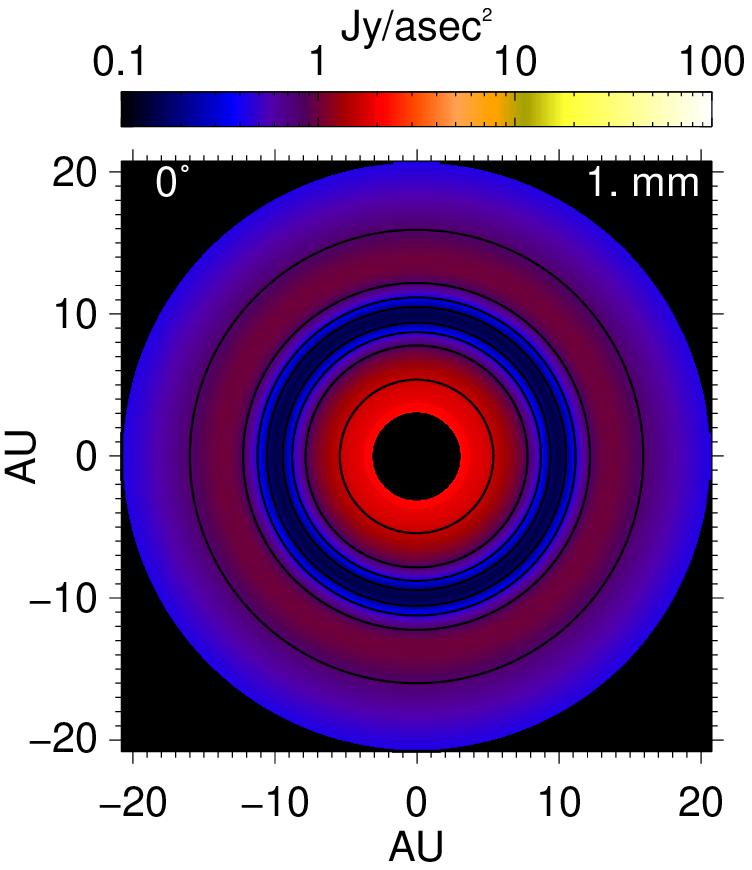}
\caption{\label{faceonlong}
Same as \figref{faceonshort}, but at 
0.1 (top), 0.3 (middle), and 1 (bottom) mm.
}
\end{figure*}

In Figures \ref{faceonshort} and \ref{faceonlong}, we show 
synthetic images of face-on disks 
without and with gaps at wavelengths 
from 1 $\mu$m to 1 mm, as indicated.  
The star is omitted in each image, but its brightness is that of 
of a 4280 K blackbody with radius $2.6\,R_{\odot}$. 
Although the apparent brightness varies as the
inverse square of the distance, the surface brightness in Jy/asec$^2$
is distance-independent. 

The top row in \figref{faceonshort} shows the scattered light images 
of the disks at 1 $\mu$m.  The 3 $\mu$m image is very similar to this 
one, just scaled to the stellar brightness at 3 $\mu$m 
and modified by the albedo according to Eq.~(\ref{eq:multscat}). 
The stellar brightness at 1 $\mu$m at a distance of 140 pc is 0.79 Jy.  
The shadow in the gap and the brightening outward of the gap 
are quite apparent.  The puffing-up of the outer edge of the gap 
also creates a further shadowing of the disk beyond the gap.  
This disk self-shadowing is also evident at longer wavelengths, 
resulting from cooling in the shadow.  
Because the disk is not well-modeled outside 20 AU, it is 
unclear whether or not the full extent of the disk beyond the gap 
rim is shadowed or not.  

The middle row in \figref{faceonshort} shows 
simulated images of the disks at 10 $\mu$m, 
where the stellar brightness is 0.055 Jy
at a distance of 140 pc.  
In the inner, warm region of the disk, the brightness is dominated 
by thermal emission from the disk surface.
The surface brightness profile is steepest at 10 $\mu$m 
because the outer disk is too cold to emit efficiently.  
At about 10 AU and beyond, the brightness profile 
becomes dominated by scattered light.  Both 
shadowing of scattered light and cooling of the disk surface 
contribute to the imaged gap structure. 

The bottom row in \figref{faceonshort} shows  
the thermal emission from the surface of the disk at 30 $\mu$m. 
The stellar brightness at this wavelength
at a distance of 140 pc is 0.0068 Jy.  As with 
the 10 $\mu$m image, the radial falloff in surface brightness 
reflects the radial temperature gradient at the disk surface.  
The gap contrast is highest at 30 microns.  This is because the 
surface temperatures of the disk are $\sim100$ K, where the blackbody 
peak is 30 microns.  Thus, observations at 30 microns will be the 
most sensitive to temperature perturbations at the surface of the disk.  
At shorter wavelengths, 
the gap contrast is driven by shadowing and illumination 
at the surface rather than actual temperature changes.  

In \figref{faceonlong}, the top, middle and bottom rows show thermal 
continuum emission at 0.1, 0.3, and 1 mm, respectively.  
At these wavelengths, the stellar emission is negligible.  
The disk becomes more optically thin toward longer wavelengths, 
so the different wavelengths probe different layers in the disk as 
well as different temperature regimes.  
In contrast to the dimple images calculated in JC09, 
the gaps are still apparent at 1 mm wavelengths because 
the shadowing effect in the gap is on a much larger scale 
than the localized dimple.  Thus, searching for gaps in disks 
with ALMA is a promising way of finding planets in during the 
planet formation epoch.  

One might expect that the shadow produced by the gap 
is offset by the brightened outer edge of the gap, creating 
a net zero effect on the spectral energy distribution (SED).  
Because the simulated region is limited in radius, a full SED
cannot be produced for the entire disk, as any emission from 
interior to 4 AU or exterior to 20 AU will be omitted.  
Nevertheless, we can assess how a gap in a disk changes the 
SED by integrating over the face on disk images.  We approximate 
the contribution from the inner 4 AU by assuming that the brightness 
profile goes as $r^{-1}$, is normalized to the brightness 
at 4 AU for the gapless disk, and is unaffected by the presence of the 
gap.  The star's emission is approximated as a blackbody 
of 4280 K and radius 2.6 $R_{\odot}$.  

\begin{figure}[htbp]
\plotone{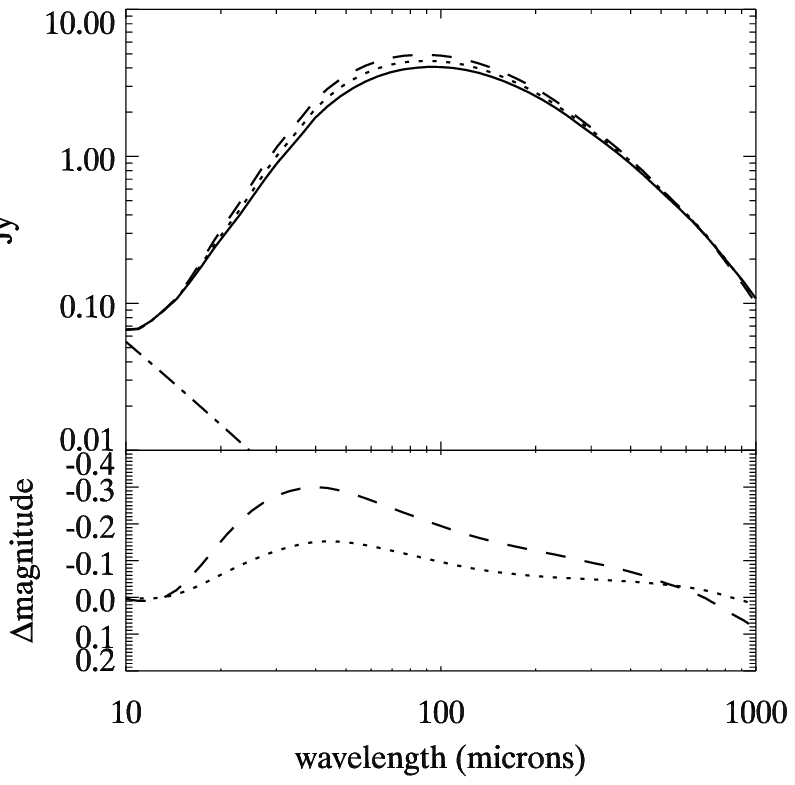}
\caption{\label{SEDdiff} SEDs of face-on disks with and without gaps
from JC models.  
The SEDs are calculated by summing the total thermal 
emission over the simulated disks.  
Top: the disk integrated spectrum of a disk without 
a gap is plotted as a solid line, while the 
disks with gaps carved by 70 and 200 $M_{\earth}$ planets are 
plotted as dotted and dashed lines, respectively.  The 
dot-dashed line shows the stellar photosphere.  
Bottom: the difference between the gapped disk models and the gap-less 
disk are plotted in magnitudes.  The 70 and 200 $M_{\earth}$ models 
are shown as dotted and dashed lines, respectively.  
The maximum deviation from the gapless model occurs at around 
$\sim40\,\mu$m, corresponding to the peak of emission for temperature 
of the disk surface at the gap radius.  
}
\end{figure}

In \figref{SEDdiff}, we show the resulting SEDs assuming that 
the system is at a distance of 140 pc.  
Since the radiative transfer calculation includes only the annulus lying 3
to 20 AU from the star, the flux approaches that of the whole disk
only at wavelengths from 10 to 100 microns, where the annulus emits
most strongly.
Nevertheless, the difference in fluxes should be representative.  
The upper panel shows that the 
gapped SEDs (dotted/dashed line for 70/200 $M_{\oplus}$)
are generally brighter than the gapless SED (solid line), 
particularly at around $30-40\,\mu$m, corresponding to the peak 
in thermal emission from the disk surface at the gap radius.  

The lower panel of \figref{SEDdiff} 
shows the difference between gapped and gapless 
SEDs in magnitudes, with negative values being brighter.  
These magnitude differences should be interpreted as upper limits 
on the amount of brightening that can occur, because emission 
from the outer disk beyond 20 AU has been omitted, and this disk 
contribution may swamp out the relative differences.  
Shortward of 10 $\mu$m, stellar flux (dot-dashed line) dominates the SED 
so the disk contribution is negligible.  
Beyond the submillimeter regime, emission from the outer disk 
becomes increasingly important, so the effect of the gap 
is relatively less than that shown.  At best, a 70 (200) $M_{\earth}$ 
planet increases the brightness of the disk by 0.15 (0.3) mag at 
44 (40) $\mu$m, or by 15 (32) \%.  
Beyond about 0.7 mm, the $\Delta$mag becomes positive for the 
200 $M_{\earth}$ gap, indicating that there is more dimming than brightening 
at radio wavelengths.  Since the maximum grain size 
assumed for calculation of the opacities was 1 mm, 
this is likely because 
at wavelengths longer than the maximum grain size in disks, the 
brightness is more representative of the dust distribution in the disk
rather than the temperature.   

In general, the conclusion is that the brightened outer rim of the gap 
more than compensates for the lack of emission in the gap 
trough.  The reason for this is the greater amount of heating 
that occurs on the outer shoulder of the gap, as described in 
\S\ref{gapstruct}.  Moreover, the contribution to the brightening 
is amplified because the surface area of the outer rim is greater 
than the surface area of the gap shadow.  However, the 
contribution of the gap to the SED is subtle, and not easily distinguished 
from a more massive disk.  The SED alone cannot be used to identify 
a partially cleared gap.  

\section{Discussion}

In JC08, we carried out similar calculations on 
local perturbations to the disk caused by less massive planets 
than the gap producing ones modeled in the present work.  
Since gaps are much larger in physical scale, they produce 
both a more significant temperature perturbation on the disk and 
a larger footprint on the disk itself.  Thus, planets that 
produce the gaps modeled in this paper are very likely to be observable, 
whereas the much more subtle perturbations modeled in 
JC08 are much harder to observe.  Imaging is necessary for 
detecting these gaps, since the effects are too subtle to 
identify in the SED alone, as described in \S\ref{sec:images}.

The width of the gaps modeled here are on the order of a few AU\@.  
If we assume that we need a resolution equivialent to 1-2 AU,
then if the disk is at the distance of Taurus, 140 pc, then the 
angular resolution required would be 0\arcsec.01, independent of 
wavelength.  Scattered light observations face the challenge of 
suppressing the starlight to a small enough working angle to 
observe the gap, in addition to the angular resolution challenge.  
ALMA should have sufficient angular resolution to resolve 
these gaps, but the question them becomes whether or not it 
will have sufficient sensitivity given the long baselines 
and sparse $u-v$ coverage required.  The size of the 
gap scales with distance, so a gap at 100 AU would be resolvable 
with 0\arcsec.1 resolution.  

We have only modeled gaps at 10 AU in this present work, but 
we can make some extrapolations to how gaps at other distances in
the disk would behave.  
The same planet mass opens a smaller gap at larger radii because 
the disk scale height increases faster than the Hill radius of the 
planet.  Thus, the temperature variations would be relatively smaller 
as the planet is moved further out.  On the other hand, 
gaps created by planets at larger radii should be more easily detectable 
because gap widths scale with distance, relaxing the 
angular resolution requirement, and the inner working 
angle is larger.  For interferometers, sensitivity also improves 
with smaller baselines.  We found that the 
greatest gap contrast was seen at the blackbody peak associated 
with the temperature of the surface of the disk.  At 10 AU,
the surface temperature is around 100 K, so the wavelength of 
greatest contrast is 30 microns.  The surface temperature is driven by 
stellar irradiation, so it should vary roughly as $r^{-1/2}$.  
This means that to some extent, the wavelength of observation may be 
tuned to the planet's orbital radius.  

\section{Conclusions}

We have calculated the thermal effects of gap-opening by planets 
in protoplanetary disks.  The thermal feedback leads to 
depression of gap troughs and puffing of gap walls, enhancing the 
observability of gaps carved by planets forming in disks.  
Planets less than one tenth of the critical gap-opening mass,  
or 10 M$_{\oplus}$ at 10 AU, do not 
create significant gaps in disks.  However, a modest gap of only 50\%,
created by a planet $\sim10\%$ of the viscous gap-opening mass or 
70 M$_{\oplus}$ at 10 AU, can induce a significant
perturbation to the temperature profile of a protoplanetary disk.
These gaps are observable in scattered light and thermal emission.  

The ability to determine the masses of planets in disks, 
together with the age of the disks and the locations 
of gaps, puts vital observational constraints on the time scales 
of planet formation.  If we find that massive planets form 
early, this might indicate that giant planets form via gravitational 
instability, which is much faster than the competing paradigm of 
core accretion.  On the other hand, if we only see gap-forming planets 
at late ages, this might indicate that core accretion is the 
dominant process.  Upper limits on the luminosity of a planet embedded 
in the disk will determine whether giant planet form more like brown dwarfs 
with a hot start \citep{2003Baraffe_hotstart} or more quiescently 
with a cold start \citep{2007Marley_coldstart}.

Temperature variations in the disk produced by shadowing and 
illumination can have profound effects on the forming planets.  
We address the consequences for Type I migration in a future paper.
These temperature perturbations can also affect
the condensation and sublimation of volatiles.  Planetesimals just
interior to the gap may become enriched in frozen volatiles, while
those just outside the gap might become depleted in volatiles via the
cold finger effect \citep{1988StevensonLunine}.  This can affect the
overall distribution of volatiles in the protoplanetary disk, shift
the snow line, and lead to enhanced planet formation in the volatile
enriched regions.

Topics for future work include carrying out radiative transfer modeling 
in a similar way on three-dimensional hydrodynamic simulations of 
gap-clearing by planets rather than relying on a simple analytic 
model.  This would allow us to capture non-axisymmetric gap characteristics, 
particularly the region immediately around the planet itself.  
Our methods can also be applied to 
the inner walls of fully-cleared inner cavities in transitional disks.  
If the inner walls are sufficiently puffed up, they can shadow the 
outer disk and create flat radial surface brightness profiles seen 
in some disks \citep{2005Grady_etal}.  
Polarized intensity images are a promising way to resolve 
protoplanetary disks \citep[e.g.~][]{Oppe08}, but geometrical effects 
make these images difficult to interpret 
\citep[e.g.~][]{HJCKuchner,2009Perrin_etal}.  
In a forthcoming paper, we will address how inclination and polarization 
affects images of disks with and without gaps.  

\acknowledgements
The authors thank an anonymous referee for constructive comments that 
greatly improved this paper.    
H.J.-C. acknowledges support for this work
through the Michelson Fellowship Program under contract with the
Jet Propulsion Laboratory (JPL) funded by NASA.  N.J.T. carried out
his part at JPL. which is managed for NASA by the California
Institute of Technology.

\bibliographystyle{apj}
\bibliography{../../../planets,../../../jang-condell,../../draft/LkCa15,gaps}

\end{document}